\documentstyle[prd,aps,preprint]{revtex}
\begin{document}
\draft

%
%

\preprint{Nisho-98/3} \title{Axionic Boson Stars in Magnetized\\ Conducting 
Media and Monochromatic Radiations} 
\author{Aiichi Iwazaki}
\address{Department of Physics, Nishogakusha University, Shonan Ohi Chiba
  277,\ Japan.} \date{June 28, 1998} \maketitle
\begin{abstract}
Axions have been argued to form
coherent axionic boson stars as well as incoherent axion gas. 
These are ones of most plausible candidates of dark matters.
Since the axionic boson stars generate oscillating electric fields 
in an external magnetic field,
they induce oscillating currents in magnetized
conducting media, which result in 
emitting radiations. We show that colliding with
a magnetic white dwarf, an axionic boson star can emit 
a monochromatic radiation with a frequency given by 
a mass of the axion.
\end{abstract}
\vskip2pc
The axion is one of most plausible candidates of dark matters
in the Universe. It is the pseudo Nambu-Goldstone boson associated with 
Peccei-Quinn U(1) symmetry\cite{PQ}, 
which was introduced to solve naturally the strong CP problem. 
The axions are produced mainly  
by the decay of axion strings\cite{kim,text} which arise when the symmetry
is broken spontaneously, 
or coherent oscillations\cite{kim,text} of axion field in the early Universe; 
the coherent oscillations 
arise when the axion gains mass through QCD instanton effects. 
These axions form incoherent axion gas 
as an axion dark matter in the present Universe.
Thus it is quite significant to confirm their existence.
Several ways of detecting these axions have been proposed.


In addition to these incoherent axions,  
the existence of coherent axionic boson stars\cite{real,iwa} 
has been argued\cite{hogan,kolb}.
It have been shown numerically\cite{kolb} that 
axion clumps are formed around the period of 
$1$ GeV owing to both the nonlinearity of an axion potential and 
the inhomogeneity of coherent axion oscillations
on the scale beyond the horizon. Their masses are about $10^{-12}M_{\odot}$
or less. Then,
these clumps contract gravitationally to axion miniclusters\cite{kolb2}
after separating out from the cosmological expansion.
They are incoherent axions localized contrary to 
axion gas distributed uniformly.
Furthermore, depending on their energy densities, 
some of these miniclusters may contract gravitationally 
to more compact coherent boson stars\cite{Tk,kolb,re}.
Eventually we expect that in the present Universe, 
there exist the coherent axionic boson stars
as well as the incoherent axion miniclusters and axion gas 
as axion dark matters. 
It has been estimated\cite{fem} that both of the axionic 
boson stars and the miniclusters may constitute a substantial fraction of 
the total mass of the axion dark matters. An observational implication
of the axion miniclusters has been discussed\cite{fem}.

In this letter we point out that colliding with the axionic boson stars,
strongly magnetized stars 
can emit detectable amount of electromagnetic radiations 
with a frequency given by the axion mass. 
As we have shown in the previous paper\cite{iwa}, 
the coherent axionic boson stars 
( we call them axion stars in this paper ) 
oscillate 
and generate oscillating 
electric fields in external magnetic fields. 
These electric fields induce oscillating electric currents
in the conducting media. Thus we expect emission of  
electromagnetic radiations in the magnetized conducting media,
when they collide with the axion star. 
Since most of them are absorbed inside the conducting media,
observable emission is made at the surface of the media.
Although the electric field itself is quite weak, 
the amount of the energy of the radiation 
is very large due to   
astrophysically large surface area of the media, e.g. white dwarfs, 
neutron stars.
Consequently detectable monochromatic radiations are expected from the media. 
Because the strength 
of the electric fields is proportional to the strength of the magnetic field,
the phenomena are revealed especially
in strongly magnetized media such as neutron stars, white dwarfs e.t.c..
We show that when the white dwarf with magnetic field $10^{6}$ Gauss 
collides with the axion star,
the luminosity of the radiations is approximately
$10^{21}$ erg/s 
$(M/10^{-14}M_{\odot})^4(m/10^{-5}\mbox{eV})^5(\sigma/10^{20}/s)$ 
where we have assumed that the radius of the white dwarf is $10^9$ cm
and that the conductivity $\sigma$ is about $10^{20}/s$ 
around the surface of the core with degenerate electrons in the white dwarf.
$M\, ( M_{\odot} )$ is the 
mass of the axion star ( the sun ) and $m$ is the mass of the axion.

Let us first explain briefly our solutions of the axion stars\cite{iwa}.
The stars are coherent axionic boson stars with masses estimated to be 
typically $10^{-12}M_{\odot}\Omega_ah^2$\cite{kolb,Tk}; 
$\Omega_a$ is the ratio of the 
axion energy density to the critical density in the Universe and 
$h$ is Hubble constant in the unit of $100$ km s$^{-1}$ Mpc$^{-1}$.
The solutions of the stars have been obtained by solving a field equation of 
the axion $a$ coupled with gravity. Since   
the axion stars are expected to have small masses, 
$10^{-12}M_{\odot}\sim 10^{-14}M_{\odot}$ for $0.01\le \Omega_ah^2\le 1$,
we have solved a free field equation of the field $a$ 
along with Einstein equations in a 
weak gravitational limit. Namely nonlinearity of axion potential
has been neglected; we found that the nonlinearity appears when masses 
of the axion stars are bigger than $10^{-9}M_{\odot}$.
It has turned out that 
our numerical solutions of the axion star with radius $R_a$ 
may be approximated by
the explicit formula,

\begin{equation}
\label{a}
a=f_{PQ}a_0\sin(mt)\exp(-r/R_a)\quad, 
\end{equation}
where $t$ ( $r$ ) is time ( radial ) coordinate and 
$f_{PQ}$ is the decay constant of the axion 
whose value is constrained 
from cosmological 
and astrophysical considerations\cite{text} such as 
$10^{10}$GeV $< f_{PQ} <$ $10^{12}$GeV.
$a_0$ is a dimensionless numerical constant
representing an amplitude of the axion field. 

We note that the solutions oscillate with the single 
frequency given by the axion mass.
In general, solutions of the axion stars with larger masses have 
many modes with various frequencies\cite{real}.  
This is a distinctive feature of the real scalar field $a$,
compared with that of boson stars\cite{re} 
composed of complex scalar fields.

The radius $R_a$ in the solutions is a free parameter and 
is related with the mass $M$ 
of the star,

\begin{equation}
\label{mass}
M=6.4\,\frac{m_{pl}^2}{m^2R_a}\quad,
\end{equation} 
with Planck mass $m_{pl}$.
Numerically, for example, 
$R_a=1.6\times10^8m_5^{-2}\mbox{cm}$ 
for $M=10^{-12}M_{\odot}$, 
$R_a=1.6\times10^{10}m_5^{-2}\mbox{cm}$ for $M=10^{-14}M_{\odot}$,
e.t.c. with $m_5\equiv m/10^{-5}\mbox{eV}$. 
A similar formula has been obtained in the case of 
the complex scalar boson stars.
We have also found an explicit relation \cite{iwa} 
between the radius and the dimensionless amplitude $a_0$ in eq(\ref{a}),

\begin{equation}
\label{a_0}
a_0=1.73\times 10^{-8} \frac{(10^8\mbox{cm})^2}{R^2}\,
\frac{10^{-5}\mbox{eV}}{m}\quad.
\end{equation}
Thus it turns out that our solutions are characterized by one 
free parameter, e.g. the mass of the axion star.
These explicit formulae are used 
for the evaluation
of the amount of radiations emitted by the axion stars in the magnetized 
conducting media.

We proceed to show how the coherent axion field of these axion stars
generates an electric field
in an external magnetic field $\vec{B}$.
The point is that the axion couples with the electromagnetic fields
in the following way,

\begin{equation}
   L_{a\gamma\gamma}=c\alpha a\vec{E}\cdot\vec{B}/f_{PQ}\pi
\label{EB}
\end{equation}
with $\alpha=1/137$, 
where  
$\vec{E}$ is electric field. 
The value of $c$ depends on the axion models\cite{DFSZ,hadron};
typically it is the order of one.  
It is easy to see how the Gauss law is modified by this interaction, 

\begin{equation}
\label{Gauss}
\vec{\partial}\vec{E}=-c\alpha \vec{\partial}(a\vec{B})/f_{PQ}\pi
+\mbox{``matter''}
\end{equation}
where the last term ``matter'' denotes an electric charge formed by 
ordinal matters. The first term in the right hand side 
represents electric charge, 
$\rho_a=-c\alpha \vec{\partial}(a\vec{B})/f_{PQ}\pi$,
formed by the axion under the magnetic field. 
Thus we find\cite{Si} that the axion star 
possesses an electric field,
$\vec{E_a}=-c\alpha a\vec{B}/f_{PQ}\pi$ associated with 
this electric charge.
Note that the strength of the field is quite 
small owing to smallness of the factor $a/f_{PQ}$; 
for instance, the factor is $\sim 10^{-8}$
for an axion star with mass $\sim 10^{-12}M_{\odot}$.

As is shown, the field $a$ oscillates in time, 
so that the electric field $\vec{E_a}$
also oscillates.
Thus electric currents induced by this field oscillate 
with the same frequency as that of the field and 
consequently electromagnetic radiations are emitted.

Here we wish to point out that there are two types of the currents induced.
One is the current, $J_m=\sigma E_a$, 
carried by ordinary charged particles, e.g.
electrons, protons etc,
which is induced by the electric field $\vec{E_a}$
in a magnetized conducting medium with a conductivity $\sigma$. 
The other one is the current, $J_a$, 
constituted by the axion field itself; since electric charge,
$\rho_a=-c\alpha \vec{\partial}(a\vec{B})/f_{PQ}\pi$, 
in eq(\ref{Gauss}) oscillates
in time, an oscillating current, 
$\vec{J_a}=-c\alpha\partial_{t}a\vec{B}/f_{PQ}\pi$ arises\cite{Si} 
associated with this 
charge owing to the conservation of the current, $\vec{\partial}\vec{J_a}-
\partial_t\rho_a=0$. 
Hence this type of the current is present even
without conducting medium as far as the magnetic field is present.

It is important to know which is dominant one in astrophysical circumstances.
The ratio of these two currents is given by 
$J_m/J_a=\sigma/m$.
Since the axion mass is constrained approximately such as 
$10^{10}/\mbox{s}<m<10^{12}/\mbox{s}$, corresponding to the above constraint
on $f_{PQ}$, the current $J_m$ is dominant in the medium with 
larger conductivity than $m$. For instance, conductivities $\sigma$ inside of 
the white dwarf or the neutron star are much larger than $m$. This is  
mainly because electrons density is
much larger than that of normal metals whose $\sigma$ is the order of 
$10^{17}$/s with room temperature. 
Thus $J_m$ dominates over $J_a$ inside of these stars with magnetic fields.
On the other hand, the current $J_a$
is dominant in the medium with smaller conductivity $\sigma \ll m$.
For instance, the conductivity around the surface of the core with 
degenerate electrons in the white dwarfs must be much smaller than $m$. 
This is because electron density at the surface must vanishes 
by the definition of the surface of the white dwarf. In a realistic 
situation there is the atmosphere of H or He with nonvanishing 
conductivity above the core of the white dwarf. The boundary between the 
atmosphere and the core is obscure; the atmosphere near the core involves
a fraction of carbons which compose the outer core and 
electrons are not fully degenerate 
in the outer core. 
Hence actual conductivity
increases with the depth in the atmosphere, 
starting at the value of $0$, but the behavior
of the value is not well understood in the region
near the boundary. Apart from the region, however, it 
reaches the large value expected 
in the degenerate crystallized core whose  
properties are well understood. As a result an average value of the 
conductivity over a region including the envelope and the outer core can be 
smaller than the mass of the axion. 
In such a case the current $J_a$ dominates over $J_m$.
Therefore we should take an appropriate current dominantly generating 
electromagnetic radiations, depending on whether or not 
the conductivity $\sigma$ is larger than $m$.

Anyway we found that the axion star 
generates the oscillating electric field $E_a$ 
under the external magnetic field and induces the 
oscillating currents $J_m$ and $J_a$ in the conducting media. 
These oscillating currents 
generate the monochromatic radiations
with the frequency given by the mass of the axion. Thus if the luminosity 
of the radiation is sufficiently large, we can determine 
the mass of the axion by the observation of the radiation. 
Thus it is important to 
estimate how large the luminosities in various media are. 
Here we take only white dwarfs
as such media, which are estimated to dominate the halo of our galaxy
and to cause the gravitational microlensing\cite{macho}. 
Thus we expect that collisions 
between the white dwarfs and the axion stars occur so frequently 
in our galaxy that these radiations can be observed.  
In addition, we assume that the mass of the axion star is 
the order of $10^{-14}M_{\odot}$ corresponding to $\Omega_ah^2\sim 0.01$:
The recent observations indicate much smaller values of $\Omega h^2$ than 
$1$; $\Omega$ is the ratio of the energy density to the critical density.

Before evaluating the luminosity we need to take into account that  
almost of all radiations generated in this mechanism are absorbed inside 
of the white dwarfs owing to the large conductivity; there are 
so many free electrons to absorb the radiations. Only radiations emitted 
around the surface of the stars can escape the white dwarfs.
Thus we need to know electromagnetic properties of the surface of the 
white dwarfs, in order to estimate the luminosity of 
the radiations emitted at the surface.
( Most of the energies of the radiations are dissipated 
in the degenerate 
cores of the white dwarfs. Hence the white dwarfs are heated and  
their temperatures increase. Among others, sufficiently 
cooled dark white dwarfs become bright again\cite{iwazaki} with 
such a collision.)

In the white dwarfs there are H or He atmospheres 
above the surface of the core with degenerate electrons.  
Opacities of the atmospheres of relatively hot white dwarfs whose 
effective temperatures are much larger than $10^4$ K, are 
known both observationally and theoretically.  
But knowledge of opacities of the 
atmospheres of the cool white dwarfs is very poor. 
In particular our concerns are 
dark white dwarfs with very low effective temperature. 
Such white dwarfs are expected 
to cause the gravitational microlensing\cite{cwd}. Their population 
is estimated to be very large; their total mass is about 
the half of the halo mass in our galaxy. 
Thus these very faint white dwarfs are dominant over ones having 
been observed.
Such white dwarfs have been cooled enough to enter a Debye cooling regime
where speed of the cooling is very fast; 
most of them in the halo must be very dim. So we do not have 
any observational evidence about their atmospheres.

Now, noting these circumstances we evaluate the luminosity of the radiations
from the surface of the magnetized conducting media. 
Since we only consider white dwarfs and the axion stars with small masses
$\sim 10^{-14}M_{\odot}$, 
the typical radius $\sim 10^9$ cm of the white dwarf is smaller than 
those of the axion stars; 
their radii are $\sim 10^{10}$ cm.  
Thus 
the white dwarfs can emit radiations when they are inside of the axion star. 
We denote a depth of a region from the surface by $d$, in which 
radiations are emitted and can escape from the white dwarf. 
We also denote an average conductivity in the region by $\sigma$.
These values are not well known so that we take them as free parameters.
Noting that only radiations from a semi-sphere facing 
observers can arrive at them, we calculate electromagnetic gauge
potentials $A_i$ of the radiations with an appropriate gauge condition,

\begin{eqnarray}
A_i&=&\frac{1}{R_0}\int_{\mbox{surface}}J_m(t-R_0+\vec{x}\cdot\vec{n})\,d^3x\\
&=&\frac{c\alpha\sigma a_0B_i}{\pi R_0}\int_{\mbox{surface}}
\sin m(t-R_0+\vec{x}\cdot\vec{n})\,d^3x\\
&=&\frac{2c\alpha\sigma a_0B_iR}{R_0m^2}\,(md\cos m(t-R_0)-
2\cos m(t-R_0+R-d/2)\sin (md/2))
\end{eqnarray}
where we have integrated it over the region around 
the surface with the depth $d\ll R$. $R_0$ is the distance 
between the observer and the white dwarf.
Here we have used the current $J_m=\sigma E_a$ with the field $a$ in 
the approximate formula eq(\ref{a}) with setting $\exp{(-r/R_a)}=1$;
the white dwarf is involved fully in the axion star so that $r/R_a\ll 1$.
On the other hand, the current $J_a$ should be used for $\sigma \le m$, 
in which case $\sigma$ should be replaced with $m$ in the above formula.
Using the gauge potentials, we evaluate the luminosity of the radiations,

\begin{equation}
L=\frac{8}{3}(\frac{\sigma}{m})^2\,c^2\,a_0^2\,B^2\,R^2\,K^2 \quad,
\end{equation}  
with $R$ being radius of the white dwarf, where 
we have taken an average both in time and the 
direction of the magnetic field.
$K^2$ is given such that 

\begin{eqnarray}
K^2&=&(m^2d^2+4\sin^2(md/2)-4md\cos(mR)\sin(md/2))/2\\
   &\cong& m^2d^2/2 \quad \mbox{for $md \gg 1$}\\
   &\cong& m^2d^2(1-\cos(mR))\quad \mbox{for $md \ll 1$}\quad.
\end{eqnarray}
In both limit $K^2$ is proportional to $m^2d^2$. Thus it turns out that
$L$ is proportional to $\sigma^2d^2$ for $m\le \sigma$, or to $m^2d^2$ for
$\sigma \le m$; as is shown soon later, the factors are large. 
We should note that the luminosity 
is proportional to the surface area $R^2$ of the white dwarf.
Thus the quantity is 
enhanced even if a luminosity per unit area in the surface is 
quite weak. This is the point we wish to stress.
Generally, phenomena caused by the axion are too faint to be detected 
owing to small factor of $m/f_{PQ}$. But in our case we have a large 
factor $R^2/m^2$ of the order of $10^{18}$. 
We also have a large factor, $\sigma^2d^2$ or
$m^2d^2$, depending on whether or not the mass is larger 
than the conductivity of the medium.   

Since it is difficult to estimate accurately the depth $d$ of a region 
in which the radiation emitted can escape from the white dwarf,
we assume for convenience that the depth is given by the penetration depth
of the electromagnetic radiation.
Then the depth is given in terms of the conductivity.
In the case that the frequency $m$ of the radiation is 
much larger than the conductivity $\sigma$, the depth $d$ is given by 
$d=1/2\pi\sigma$. On the other hand, the depth is given by 
$d=\sqrt{1/2\pi \sigma m}$ for $m\ll \sigma$. 
Hence it follows that the factor $\sigma^2d^2$ of the enhancement 
is the order of $\sigma /m$ for $m\ll \sigma$ and 
the factor $m^2d^2$ is the order of $m^2/\sigma^2$ for $\sigma \ll m$.  
They are much larger than $1$.

Then with the assumption we find that the luminosity is numerically given by 

\begin{eqnarray}
\label{l1}
L&\sim&10^{21}\mbox{erg/s}
\,B_6^2\,R_9^2\,\sigma_{20}\,m_5^5\,M_{14}^4\,c^2 \quad 
\mbox{for $m\ll \sigma$}\\
&\sim&10^{21}\mbox{erg/s}
\,B_6^2\,R_9^2\,m_5^8\,M_{14}^4\,c^2/\sigma_5^2 \quad 
\mbox{for $m\gg \sigma$}
\label{l2}
\end{eqnarray}
where 

\begin{equation}
B_6=\frac{B}{10^6\mbox{G}},\quad R_9=\frac{R}{10^9\mbox{cm}},\quad
\sigma_{20}=\frac{\sigma}{10^{20}/\mbox{s}},\quad
\sigma_{5}=\frac{\sigma}{10^{5}/\mbox{s}},\quad
M_{14}=\frac{M}{10^{-14}M_{\odot}}\quad.
\end{equation}  
We have used the penetration depth $d=1/2\pi\sigma$ for $m\gg \sigma$, and 
$d=\sqrt{1/2\pi \sigma m}$ for $m\ll \sigma$, respectively. In both case 
we have simply assumed that 
values of both dielectric constant and magnetic permeability 
are the same as those of the vacuum. Numerically, 
$d\sim 10^{5}$ cm$/\sigma_5$ for $m\ll \sigma$ and
$d\sim10^{-5}$ cm $(\sigma_{20}m_5)^{-0.5}$ for $m\ll \sigma$.

$L$ in eq(\ref{l1}) is the luminosity of the radiation generated by the 
ordinary current $J_m$ dominant in the medium with $\sigma \gg m$.
This is the case that the atmosphere of the white dwarf has a vanishing 
conductivity and the core with degenerate electrons has much large 
conductivity, e.g. $10^{20}/\mbox{s}$, even at the surface of the core.
This is the case of the sharp boundary of the white dwarfs.
On the other hand, $L$ in eq(\ref{l2}) is the luminosity of the radiation 
generated by the current $J_a$ dominant in the medium with $\sigma \ll m$.
This is the more realistic case that a small conductivity 
of the atmosphere increases with the depth and reaches 
a large value at the core. 
Then, the average value of the conductivity in the 
region with the depth $d$ may be much smaller than $m$; for instance,
$\sigma\sim 10^5/\mbox{s}$. 

We would like to mention that the result in eq(\ref{l2}) heavily depends 
on the mass of the axion. If the mass is given by $10^{-4}$ eV, 
the luminosity becomes large such as $L\sim 10^{29}$ erg/s. 
Similarly, the above results depend on the unknown parameters, mass $M$ of 
the axion star, magnetic field $B$, conductivity $\sigma$,
or depth $d$ of the white dwarf.
Hence it is difficult to obtain the accurate value of the luminosity.
But it is interesting that there is a possible range of 
the parameters for us to be able to observe the radiations.

The radiations are emitted during the collision between the axion star 
and the white dwarf. Thus the emission continues approximately for 
$2\times10^3$ sec which is taken for one passing the other,
when the relative velocity is $10^{-3}c$; the velocity 
is supposed to be typical one of matters composing the halo.
On the other hand if the axion star is trapped gravitationally by the
white dwarf, the emission goes on longer.  

In the above evaluation we have assumed implicitly that 
the axion star colliding with the white dwarf does not received 
any effects from it.
Since the mass of the axion star is much smaller than typical mass 
$\sim 0.5\times M_{\odot}$ of the white 
dwarf, the form of the coherent axionic boson star may be fairly deformed by 
the gravitational effect of the white dwarf. Especially the density of 
a part of the axion star occupied by the white dwarf will increase. Then the 
amplitude $a_0$ around the part becomes so large that
the luminosity is enhanced. These effects of the deformation and 
the dissipation of the energy will result 
in the axion star being trapped by the white dwarf.

Finally we point out that the rate of the collision between the axion star 
and the white dwarf in our galaxy is large to be detectable,
if the half of the halo is composed of the axion stars and the other 
half is composed of the white dwarfs. Suppose that the distribution 
of the halo with total mass $\sim 4\times 10^{11}M_{\odot}$ is given 
such that its density $\propto(r^2+3R_c^2)/(r^2+R_c^2)^2$ with 
$R_c=2\sim 8$ kpc where $r$ denotes a radial coordinate with the origin 
of the center of the galaxy. Then it is straightforward to evaluate the rate
of the collisions in a solid angle $\omega$ per year,

\begin{equation}
0.5/\mbox{year}\times\frac{1}{M_{14}^3}\,
\frac{1}{m_5^4}\,\frac{\omega}{5^{\circ}\times5^{\circ}}
\end{equation}  
where relative velocity between the axion star and the white dwarf is 
assumed to be $10^{-3}c$. We have taken into account the fact that the 
earth is located at about $8$ kpc from the center of our galaxy, simply by 
counting the number of the collisions arising in the region 
from $8$ kpc to $50$ kpc. The rate is not necessarily large enough, 
but not so small for us not to be able to detect the radiation.

In summary, we have shown that the electromagnetic radiations are emitted
with the collision between the white dwarf and the axion star.
The radiations are monochromatic with the frequency given 
by the mass of the axion. Although the luminosity heavily depends 
on the unknown parameters, e.g. mass of the axion,
there is a range of the parameters with which the luminosity is 
sufficiently large to be observable.
Therefore, the detection of the radiations makes us determine 
the value of the mass. 
Both monochromatic radiations discussed in this paper and thermal
radiations caused by heating up dark white dwarfs with the collision
discussed in a previous paper\cite{iwazaki} can be used 
for the detection of the coherent axionic dark matters.

The author would like to express his thank for useful discussions 
and comments to 
Professors M. Kawasaki and R. Nishi, and also for
the hospitality in Tanashi KEK. This research is supported by 
the Grant-in-Aid for Scientific Research from the Ministry of Education,
Science, Culture and Sports No. 10640284





\end{document}